\documentstyle[prl,aps,graphicx,multicol]{revtex}
\input psfig.sty

\begin{document}

\title{Origin of heavy quasiparticles in UPt$_3$}

\author{G. Zwicknagl~$^1$, A.~N.~Yaresko$^{2}$ and P.~Fulde$^3$} 
\address{$^{1}$ Institut f\"ur Mathematische Physik,
Technische Universit\"at Braunschweig, Mendelssohnstra\ss e 3, D-38106
Braunschweig, Germany}
\address{$^{2}$ Max Planck Institut f\"ur Chemische Physik fester Stoffe,
N\"othnitzer Stra\ss e, 40, D-01187 Dresden, Germany}
\address{$^{3}$ Max Planck Institut f\"ur Physik komplexer Systeme,
N\"othnitzer Stra\ss e, 38, D-01187 Dresden, Germany}

\date{\today}
\maketitle

\begin{abstract}
We propose a microscopic description of heavy fermions in UPt$_3$. It is based
on the assumption that two of the three 5$f$ electrons of the U ions are
localized. Band-structure calculations based on this supposition reproduce
the observed de Haas-van Alphen frequencies very well. The observed 
enhancement of the
quasiparticle mass as compared with the LDA band mass results from the local
Coulomb interaction of the delocalized $f$ electrons with the localized
ones. Diagonalization of the Coulomb interaction matrix yields the 
level scheme of the
localized $f^2$ states. Assuming a splitting of the ground-state doublet by the
crystal field of similar size as in UPd$_2$Al$_3$ results in a mass enhancement
factor of order 10 in agreement with experiments.
\end{abstract}

\vspace{1cm}

\noindent PACS.~~~~~~~~
\begin{minipage}[t]{15cm}
71.10.-w, 71.10.Ay, 71.18.+y, 71.27.+a
\end{minipage}
\vspace{1cm}

\newcommand{\gsim}{\mathrel{\raise.3ex\hbox{$>$\kern-.75em\lower1ex\hbox{$\sim$}}}}
\newcommand{\lsim}{\mathrel{\raise.3ex\hbox{$<$\kern-.75em\lower1ex\hbox{$\sim$}}}}

\begin{multicols}{2}
\narrowtext

\vspace{1cm}

The intermetallic compound UPt$_3$ is a well known heavy fermion system
\cite{StewartUPt3PRL,Stewartreview}. The Sommerfeld $\gamma$ coefficient of the linear low-temperature
specific heat is strongly enhanced, i.e., 
$\gamma$ = 420 mJ/(mole\ U $\cdot$K$^2$)
and so is the Pauli-like spin susceptibility $\chi_S$. 
Both findings
can be explained by attaching a large effective mass m$^*$ to the
quasiparticles, i.e., one which is by a factor of order 20 bigger 
than the band mass m$_b$ obtained from local density approximation
(LDA) calculations for the electronic structure
\cite{ABCUPt,Normanetal}. Indeed,
heavy quasiparticles have been observed in de Haas-van Alphen (dH-vA)
experiments 
\cite{TailleferLonzarich,Tailleferetal,KimuraUPt33}. These experiments
unambiguously confirm that UPt$_3$ has to be regarded as a strongly
correlated Fermi liquid. Although a complete picture of the low
temperature phase of UPt$_3$ has emerged, a comprehensive theoretical picture 
of the heavy quasiparticles is still missing. It is generally accepted
that the latter are derived from the U 5f states. The number of
itinerant U 5f electrons as well as the microscopic mechanism yielding
the high effective masses are still controversial. 
It has been considered a success of the LDA that the dH-vA frequencies 
could be related to extremal orbits on the
Fermi surface obtained by band-structure calculation which treats the
U 5f states as itinerant. From these findings, however, one should not
conclude that the U 5f states are ordinary band states which can be
described by conventional electronic structure calculation.
As pointed out above, the
calculated energy bands are too broad for explaining the effective
masses. Moreover, they are too small in order to fit the photoemission
data \cite{JWAllenPESReview}. The
latter shows a broad peak just below the Fermi energy E$_{\rm F}$. 
It is quite different from the data of heavy- fermion systems
involving Ce$^{3+}$ instead of
U ions, such as CePt$_3$. Here a small Kondo resonance is observed near
E$_{\rm F}$ together with a broad 4$f$ peak approximately 2 eV below that
energy\cite{BaerPESCePt,RBPES}. We add that the photoemission 
data near E$_{\rm F}$ of UPt$_3$
resembles that of UBe$_{13}$ and UPd$_2$Al$_3$, two other heavy-fermion
systems. It is also similar to that of UPd$_3$, which has localized 
5$f$ electrons except that in this case there is no $f$ weight left 
at E$_{\rm F}$. From the above it
follows that a more microscopic understanding of the heavy-quasiparticle
excitations is highly desirable. It seems clear that the Kondo effect as well
as more exotic versions of it \cite{ExoticKondo} can be excluded for 
UPt$_3$, for various reasons.

There are indications that U ions may have two types of $f$ orbitals, i. e.,
with localized and with delocalized electrons, depending on the degree of
hybridization. For example, quantum chemical calculations for uranocene
U(C$_8$H$_8$)$_2$ have demonstrated this \cite{lanthanocenes}. They have shown
that the system has low-lying excitations which are due to local rearrangements
of the 5$f$ electrons. Also neutron inelastic scattering experiments on
UPd$_2$Al$_3$ point towards the existence of localized as well as delocalized
5$f$ electrons in that material \cite{UPd2Al3Nature}. Note that the
susceptibilities of UPt$_3$ and UPd$_2$Al$_3$ look very much alike. These
findings are the basis of our theory.

The aim of the present letter is to suggest
that in UPt$_3$ we deal with localized as well as delocalized 5$f$ electrons
and that the coupling of the two subsystems results in heavy
fermions. The mass enhancements of the
quasiparticles are shown to follow from the local exchange interaction of the
delocalized 5$f$ electrons with localized f$^2$ configurations. 
The situation resembles that in Pr metal  where a  mass enhancement of the 
conduction electrons by a
factor of 5 results from virtual crystal field (CEF) excitations of localized
4$f^2$ electrons \cite{WhiteFulde}. As shown below, the theory suggested 
here allows for an
explanation of the observed dH-vA frequencies. The attained accuracy is
similar, if not better than that of previous LDA band structure calculations,
although the Fermi surface shows distinct differences in the two cases.

In the following we consider two of the 5$f$ electrons as localized, in
agreement with the absence of any Kramers doublets in cases where a CEF
splitting of U states has been observed. We put the localized $f$ electrons
into 5$f$ j=$\frac{5}{2}$ orbitals, with j$_{\rm z} =\pm \frac{5}{2}$ and $\pm
\frac{1}{2}$. This choice is also consistent with the 
observed ground state, e.g., of UPd$_2$Al$_3$ 
\cite{UPd2Al3CEFICPTM,UPd2Al3CEFNeutron}.
The 5$f$ j=$\frac{5}{2}$, j$_{\rm z} =\pm
\frac{3}{2}$ states, on the other hand, are treated as itinerant 
band electrons. This yields the observed dH-vA
frequencies when subsequently a band structure calculation is done. 

\begin{figure}[t b]
\begin{center}
\includegraphics[angle=270,width=8.5cm]{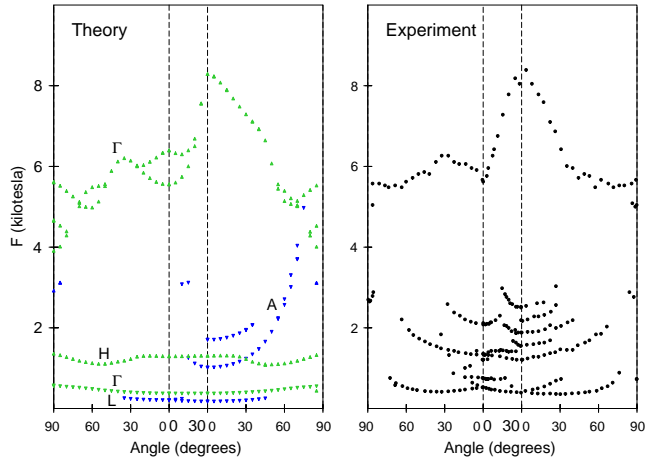}
\end{center}
\vspace{0.5cm}
\caption{DeHaas-vanAlphen cross sections as calculated within the
present theory and from experiment (Kimura et al).
Green and blue triangles correspond to the first and second band, 
respectively.
The labelling indicates the origin of the orbit. Branches without
labels are derived from those extremal areas of the lower band which are
not centered on symmetry points.}
\label{fig:CrossSections}
\end{figure}

A plot of the calculated dH-vA frequencies and of the observed ones is
shown in Figure \ref{fig:CrossSections}. We should like to emphasize that
the theoretical data were obtained without any adjustments of the
bands.

\begin{figure}
\hspace{-0.5cm}
\includegraphics[width=4.2cm]{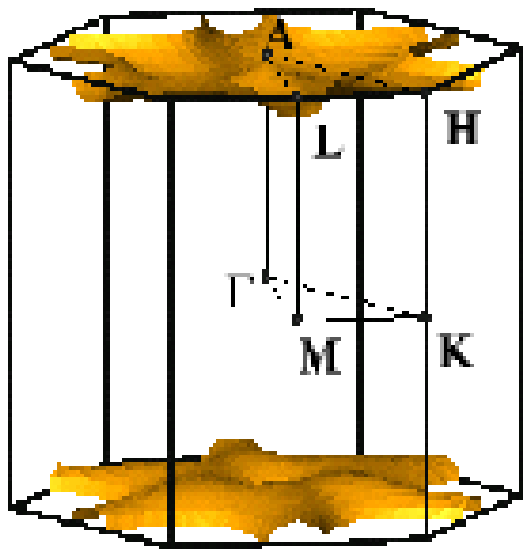}\hspace{-0.6cm}
\includegraphics[width=5.6cm]{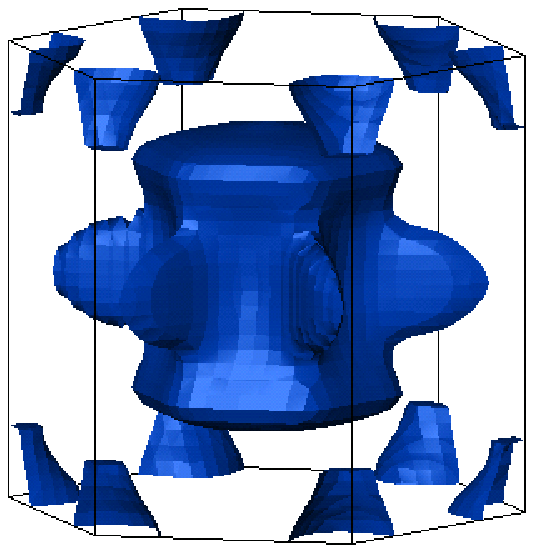}
\caption{Fermi surfaces calculated assuming that one of the uranium 
5f-electrons (j=5/2, j$_{\rm z}$=$\pm$3/2) is 
included in the Fermi surface volume while the remaining two are treated as 
localized. Two bands are contributing to the the Fermi surface.
Calculated Fermi surface (a) from band 1, (b) from band 2. The small 
$\Gamma$-centered hole surface derived from band 2 is not displayed
here. The symmetry points of the Brillouin zone are indicated in (a)}
\label{fig:FermiSurface}
\end{figure}

The Fermi surface from which the data are derived is displayed in
Figure 
\ref{fig:FermiSurface}. 
It is formed by two bands which are doubly degenerate and which are
derived from the  5$f$ j=$\frac{5}{2}$, j$_{\rm z} =\pm
\frac{3}{2}$ states.
The first band gives 
rise to a pancake-like hole surface which is centered at A. The surfaces in 
neighboring cells are connected by arms along AL. These features imply 
the existence of open orbits for magnetic fields pointing along the 
a-direction as suggested by magnetoresistance
data\cite{LonzarichNew}. The second band intersects the
Fermi surface several times giving rise to a $\Gamma$-centered hole 
ellipsoid which is enclosed by a large closed $\Gamma$-centered 
electron surface. The latter is strongly anisotropic in the basal
plane. Finally, there are closed ellipsoids centered at H. The
assignment of the observed quantum oscillation frequencies to
extremal orbits on this Fermi surface is as follows:
The thermodynamically most important orbit $\omega$ is assigned to the 
$\Gamma$-centered strongly anisotropic electron surface. The variation
with magnetic field direction of the dH-vA frequencies is
quantitatively reproduced. Of particular interest is the spitting
obtained in the a-c plane. This feature which is also found in LDA
calculations has been confirmed by recent experiments
\cite{LonzarichNew}.

The observed effective masses and calculated band masses on this 
sheet are listed in Table \ref{tab:eff_masses} for selected directions of the magnetic
field. 
\begin{table}
\caption{Comparison of the calculated band masses $m_b$ 
and measured effective mass over the bare electron mass $m_0$ for 
different directions of the magnetic field.} 

\vspace{0.5cm}
\label{tab:eff_masses}
\begin{tabular}{l c c c}
 & $\frac{m_b}{m_0}$ & $\frac{m_b}{m_0}$ & $\frac{m^*}{m_0}$ \\[2mm]
 & (LDA) \cite{KimuraUPt33} &  (present theory) &  (exp)
 \cite{LonzarichNew} 
\\[2mm]
\hline
H $\parallel$ c axis & 5.09 & 17 & 110 \\
H $\parallel$ a axis & 5.79 & 6 and 9.5 & 82 \\
H $\parallel$ b axis & 6.84 & 14 & 94\\
\end{tabular}
\end{table} 
The enhancement of the observed effective masses m$^*$ over the
ones calculated within the present theory m$_b$ is of the order of
10. It can be quantitatively explained by the interaction of the
itinerant 5$f$-states with the localized f$^2$ configuration. The
ratio $\frac{m^*}{m_b}$ is obtained in analogy to Ref. 
\cite{WhiteFulde} resulting in the equation

\begin{equation}
\label{1}\frac{m^*}{m_b} = 1 + 4 a^2 N(0) \frac{2 |M|^2}{{\tilde \delta}}~~~. 
\label{eq:effmass}
\end{equation}

Here N(0) is the density of states at E$_{\rm F}$ as obtained from our
band structure when two 5$f$ electrons are kept localized. The prefactor a is
the 5$f$-weight per spin and U atom of the conduction electron states 
near E$_{\rm F}$. The matrix element M describes the transition
between the ground-state
singlet and the low-lying excited singlet state at energy 
${\tilde \delta}$ of the localized $f^2$ subshell in the presence of a CEF. 
With appropriate numbers put into Eq. (\ref{1}) we find $m^*/m_b \simeq$ 10.

In the following, we present the details of the calculations.
Band-structure calculations have been performed starting from the 
self-consistent LDA potentials but excluding the 
U 5$f$~ j=$\frac{5}{2}$, j$_z$=$\pm \frac{5}{2}$ and j$_z$=$\pm
\frac{1}{2}$ states from forming bands. The localized 5$f$ orbitals 
are accounted for in the self-consistent density and, concomitantly,
in the potential seen by the conduction electrons. The 5$f$-bands are 
calculated by solving the Dirac equation  \cite{ABCUPt}. The intrinsic
bandwidth of the itinerant U 5$f$ j=$\frac{5}{2}$, j$_z$=$\pm\frac{3}{2}$ is
taken from the LDA calculation while the position of the
corresponding band-center C is chosen such that the density
distribution of the conduction states as obtained within LDA remains 
unchanged. The position of the $f$ band relative to the calculated Pt d states 
is consistent with photoemission data. The band centers of the
remaining U 5$f$ states are shifted to higher energies. As a
consequence the corresponding orbitals cannot form bands in the energy 
range of interest. They affect the bands in the vicinity of the Fermi 
level via the hybridization tails only. It was found that the U 5$f$
bands with j$_{\rm z} = \pm \frac{3}{2}$ hybridize
strongly near the Fermi level so that the $f$  occupancy per U atom for the
delocalized 5$f$ electrons amounts to n$_{\rm f}$ = 0.65 indicating
that we are dealing with a mixed valent situation.

We next turn to the discussion of the localized U 5$f$ states. 
The multiplet structure of the localized $f^2$ states is calculated by
diagonalizing a 6 x 6 Coulomb matrix. The spin-orbit splitting is rather large
and therefore a jj-coupling scheme is used. This simplification gives six
2-particle states build from 
$|j = \frac{5}{2}, j_z=\pm \frac{5}{2} \rangle$ and
$|j = \frac{5}{2}, j_z=\pm \frac{1}{2} \rangle$. The resulting
eigenstates are
generally no longer eigenstates of the total angular momentum J$^2$, 
but remain eigenstates of J$_{\rm z}$. The Coulomb matrix elements are 
calculated following Condon and Shortley\cite{CondonShortley}. Inputs are the 
Slater-Condon parameters F$^{\rm
K}$ (Coulomb integrals) and G$^{\rm K}$ (exchange integrals) which 
are evaluated with the
radial function R$^{\rm U}_{{\rm f},
\frac{5}{2}} ({\rm r})$ for U which is determined from a 
self-consistent band structure
potential. The radial function is calculated for an energy given 
by the center of gravity of the 5$f$ bands. The required 
integrations are done within the
atomic sphere surrounding the U ion. Diagonalization of the matrix 
yields a doubly degenerate ground state J$_{\rm z} = \pm 3$ which must be an 
eigenstate of J = 4. Note that the Pauli principle permits even values of 
J only, i.e., J = 0, 2,
4 in our case. The states 
$|j= \frac{5}{2}, j=\frac{5}{2}; {\rm J} = 4, J_{\rm z} =
\pm 3 \rangle$ have an overlap of 0.865 with the Hund's rule ground state
$^3H_4$ derived from the LS-coupling scheme. Therefore the choice of jj vs. LS
coupling should only weakly affect the results obtained for the ground-state
multiplet. The two-fold degeneracy of the ground-state is lifted by a CEF
yielding the two states

\begin{eqnarray}
\label{2}| \Gamma_3 \rangle & = & \frac{1}{\sqrt{2}} (|J = 4 ; J_z = 3 \rangle
+ | J = 4; J_z = -3 \rangle )\nonumber \\
| \Gamma_4 \rangle & = & \frac{1}{\sqrt{2}} (|J = 4 ; J_z = 3 \rangle - |
J = 4; J_z = -3 \rangle )~~~. 
\end{eqnarray}

Note that $|\Gamma_4 \rangle$ has been suggested as ground state of
UPd$_2$Al$_3$ \cite{UPd2Al3CEF,UPd2Al3CEFICPTM}. We assume that the 
splitting energy ${\tilde \delta}$ between 
$|\Gamma_4 \rangle$ and $| \Gamma_3 \rangle$ is of order 20 meV as in
UPd$_2$Al$_3$ \cite{UPd2Al3CEFNeutron}. The next-higher eigenstates of
the Coulomb matrix is a doublet with J$_z=\pm$ 2. The excitation
energy separating it from the ground state is rather large (0.4 eV). 
We therefore neglect all higher levels. For the evaluation of the 
effective mass enhancement we need to determine the coupling between 
the localized and delocalized $f$ electrons. It is directly obtained
from the expectation values of the Coulomb interaction U$_{\rm Coul}$ 
in the 5$f^3$ states. For
the latter we use the product states 
$| f^2; J = 4, J_z = \pm 3 \rangle \otimes |f^1; j
= \frac{5}{2}, j_z = \pm \frac{3}{2} \rangle$. The difference  $\langle f^1;
\frac{5}{2}, \frac{3}{2} | \otimes \langle f^2; 4, 3~ | U_{\rm Coul} |~ f^2; 4,
3 \rangle \otimes |f^1; \frac{5}{2}, \frac{3}{2} \rangle - \langle f^1;
\frac{5}{2}, \frac{3}{2} | \otimes \langle f^2; 4, -3~ | U_{\rm Coul} |~ f^2;
4, -3 \rangle \otimes | f^1, \frac{5}{2}, 
\frac{3}{2}\rangle$ amounts to -0.38 eV from which we deduce the transition
matrix element 

\begin{equation}
\label{3}M = \langle f^1; \frac{5}{2}, \frac{3}{2} |\otimes \langle \Gamma_4~
| U_{\rm Coul} |~\Gamma_3 \rangle \otimes |f^1; \frac{5}{2}, \frac{3}{2}
  \rangle  = 0.19 eV~~~.  
\end{equation}

We may also rewrite M in form of an exchange coupling between the
itinerant and localized $f$ states. In that case the Land\'e $g$
factor 
for the
localized $f^2$ states must be determined. It turns out that this 
$g$ factor is
$g_{\rm eff} = 0.63$ and that the exchange integral is $I \simeq 1 eV$, which
is of the correct size for 5$f$ electrons. 

Let us now turn to the mass enhancement Eq. \ref{eq:effmass}. The density 
of states N(0) corresponding to the Fermi surface in 
Fig. \ref{fig:FermiSurface} 
equals $\simeq $ 15.5 states/ (eV cell), the other parameters are 
$|$M$|^2$ = 0.036 eV$^2$, 4a$^2$=0.13 and 
${\tilde \delta} \simeq$ 20 meV. 
When these parameters are put into Eq. (\ref{1}) a mass enhancement 
$ \frac{m^*}{m_b}$ of approximately 8.3 is obtained. This is in
excellent agreement with experiment.

The above scenario provides for a satisfactory explanation of the heavy
quasiparticles observed in UPt$_3$. The different hybridization behavior of the various U 5f orbitals requires further investigations.
In a separate study we shall investigate the effect of strong local electron
correlations on the different LDA hybridization matrix elements.
 In
concluding we want to point out that the way of treating the $f$ electrons in
UPt$_3$ may be also applicable to other uranium compounds.
\vspace{1cm}

{\section*{\bf Acknowledgement}}

We would like to thank M. Kagan and A. Y. Perlov for participating at the
beginning stage of this work. We want also to thank J. W. Allen,
N. E. Christensen and P. Thalmeier for a number of helpful discussions and 
B.Yu. Yavorsky for providing us with the dHvA program. We are grateful
to S. R. Julian and G. G. Lonzarich for making their experimental data
available to us prior to publication.




\end{multicols}

\end{document}